\documentclass{emulateapj}

\newcommand{\lya}{Ly$\alpha$}

\newcommand{\msun}{$M_{\odot}\;$}

\newcommand{\msunperyear}{$M_{\odot} \; yr^{-1}$}

\slugcomment{}

\shorttitle{A Physical Model of \lya\ Emitters}
\shortauthors{Tilvi et al.}

\begin{document}

\title{A Physical Model of \lya\ Emitters}

\author{Vithal Tilvi\altaffilmark{1}, Sangeeta Malhotra\altaffilmark{1},  
James E. Rhoads\altaffilmark{1}, Evan Scannapieco\altaffilmark{1},
Robert J. Thacker\altaffilmark{2},
Ilian T. Iliev\altaffilmark{3,4}, and Garrelt Mellema\altaffilmark{5}}

\altaffiltext{1}{School of Earth and Space Exploration, Arizona State University,  Tempe, AZ 85287 ; tilvi@asu.edu}
\altaffiltext{2}{Saint Mary's University, Halifax, NS, Canada, B3H 3C3.}
\altaffiltext{3}{Universitaet Zuerich, Institut fuer Theoretische Physik, CH-8057 Zuerich, Switzerland.}
\altaffiltext{4}{Astronomy Centre, Department of Physics \& Astronomy, Pevensey II Building, University of Sussex, 
			Falmer, Brighton BN1 9QH, United Kingdom.}
\altaffiltext{5}{Dept. of Astronomy and Oskar Klein Centre, AlbaNova, Stockholm University, SE-10691 Stockholm, 
			Sweden}

\begin{abstract}
We present a simple physical model for populating dark matter halos with \lya\ emitters (LAEs) and 
predict the properties of LAEs at$z\approx 3-7$. 
The central tenet of this model is that the \lya\   luminosity is proportional to the star formation rate (SFR) 
which is directly related to the halo mass accretion rate. 
The only free parameter in our model is then the star formation efficiency (SFE). 
An efficiency of 2.5$\%$ provides the best fit to the \lya\  luminosity function (LF) at redshift $z = 3.1$, 
and we use this SFE to construct \lya\  LFs at other redshifts. 
Our model reproduces the \lya\ LFs, stellar ages, SFR $\approx 1-10\; M_{\odot}$ yr$^{-1}$, 
stellar masses $\sim  10^{7}$  to $10^{8}$ \msun,
 and the clustering properties of LAEs at $z\approx 3-7$. 
 We find the spatial correlation lengths $r_{o} \approx 3-6$ $h^{-1}$ Mpc, in agreement with the observations. 
 Finally, we estimate the field-to-field variation $\approx $30$\%$ for current volume and flux limited surveys, again consistent with observations. 
 Our results suggest that the star formation, and hence \lya\ emission in LAEs can be powered by accretion of new material. Relating the accreted mass, rather than the total mass, to the \lya\ luminosity of LAEs naturally gives rise to their duty cycle.
  \end{abstract}


\section{Introduction}

\lya\  emitting galaxies (LAEs) are selected on the basis of strong \lya\  emission line, irrespective of other galaxy 
properties  \citep[e.g.][]{cow98, rh00,rho01,rh04, fyn01,  aji03, mat05, ta05,wan05, ga06, shi06, tap06, mur07, 
nil07, ouc08}.
 However, high-redshift galaxies selected on the basis of this one property are reasonably uniform in some of their 
 other properties\footnote {In this  paper we restrict 
our   studies to only compact LAEs (at $z=3-6.6$) detected using narrow-band excess. 
Other \lya\ emitting objects (e.g. \lya\ blobs and   AGNs) are typically much  more 
energetic and are probably fueled by   AGN activity.}. 
 For example, the inferred stellar mass of LAEs at $z <$ 5 is typically small, $\sim10^{6}$  to$10^{9}$ \msun
 \citep{ga06,pi07, fi07,pen09} and they often have large \lya\ equivalent width (EW) indicating a young stellar population 
\citep{mr02},  which is also supported by the blue color of these galaxies 
  \citep{ve05, fi07, fi08,ga06, pi07}   especially at high redshifts. 
  Active galactic nuclei (AGNs) are ruled out as sources of strong \lya\ emission in LAEs due to non-detection of X-ray emission 
   \citep{ma03,  wa04, ga07}  and the lack of high ionization lines in the optical spectra 
  \citep{da04, da07, wan09}.
  LAEs have moderate SFRs $\approx$  5-8 \msunperyear\   \citep[e.g.][]{pi07, ta05} 
and spatial correlation lengths of  $\approx 3-5 h^{-1}\; \rm Mpc$, albeit with a substantial uncertainty 
\citep{ouc03, ko07, ga07}.

Despite the increasing number of LAE observations, theoretical understanding of LAEs is still in early 
stages, primarily due to a poor understanding of physical properties including star formation, 
stellar initial mass function, \lya\  escape fraction, and the duty cycle of the \lya\  phase. 
There have been several theoretical studies of LAEs based on cosmological simulations 
\citep[e.g.][]{ba04,dav06,tas06, shi07,na08},  semi-analytical models  \citep[e.g.][]{le06,  kob07,kob09,
day08, sam09}
 and analytical models 
  \citep[e.g.][]{ha99,di07, ma07,  st07, fe08}
 that relate the total halo mass to the \lya\  luminosity of LAEs. 
 Such a linear relationship between the halo mass and \lya\  luminosity often leads to an overprediction of the number 
 density of LAEs. 
 To reconcile the mass distribution of halos to the luminosity function (LF), one needs to either assume a 
 small escape fraction of \lya\  photons (which fails to account for the large \lya\  EWs observed) or introduce a 
 duty cycle 
 \citep[e.g.][]{st07, na08} 
 which adds another parameter to the models. 
 In addition, the complexity and large number of variable parameters in many models motivate the 
 development of a simple approach, which is particularly useful in understanding the nature of LAEs 
 observed at high redshifts.

In this paper, we present a physical model to populate dark matter (DM) halos with LAEs in a cosmological 
simulation, and predict the abundances and physical properties at $z\approx$ 3-7. 
This model differs fundamentally
 from many of the earlier studies in that we relate mass accreted, as opposed to the total halo mass, to the 
 \lya\  luminosity. 
 Mass accretion onto halos via smooth infall and accretion due to mergers of a specific mass 
 ratios has been shown to have distinctly different clustering behavior 
 \citep{sca03}.
 However, in our current work we do not distinguish between smooth accretion and the accretion due to 
 mergers. 
 In other words, in our model the LAEs are undergoing an episode of star formation driven by 
 accretion of fresh material onto the halos, independent of whether the accretion is due to mergers or via 
 smooth infall. 
 While there is no direct observational evidence showing a relation between the baryons 
 accreted and the  \lya\  luminosity, recent studies 
  \citep[e.g.][]{dek09,ker09} 
have shown  that such cold accretion of new material can drive star formation in galaxies.

Using the Millennium simulation, 
\citet{gen08} 
found that the high SFRs observed in$ z\approx$ 2 galaxies can be explained by continuous mass accretion. 
Similar studies 
 \citep[e.g.][]{hai00, far01}
 have shown that the baryons inside high-redshift halos can release significant amount of gravitational binding energy 
 in the form of \lya\ luminosity as the baryons condense within DM potential wells. 
 The \lya\ emission resulting from this mechanism would, however, lead to low surface brightness extended \lya\ emitters or \lya\  blobs which are more diffuse than LAEs.

In this model, we assume that LAEs do not contain large amounts of dust, and hence most of the 
hydrogen ionizing photons will be absorbed, while most of the \lya\ photons will escape 
 \citep{ga06, kob07}. 
These assumptions are needed to produce large 
EWs of \lya\  line 
\citep{mr02}. 
It has also been shown that the velocity gradients in the 
gas can facilitate the \lya\ photon escape and making them less susceptible to dust absorption
 \citep{dij06}. 
In addition, \lya\  photons can preferentially escape from LAEs if the dust is 
 primarily in cold, neutral clouds 
 \citep{ha99,ha06, fi08}. 
 Our assumption of large escape fraction of \lya\  photons naturally yields large \lya\ EWs even without 
 appealing to metal-free Population III stars, whose contributions are constrained by non-detection of 
 the He II (1640) line 
 \citep{da04, da07, no08, wan09}. 
In such conditions, 
 the \lya\  line becomes a direct measure of the SFR, which is proportional to the accretion of fresh 
 gas onto the galaxy. The constant of proportionality (the star formation efficiency, SFE) between 
 accretion rates and \lya\ luminosity is the only free parameter in our model.

This paper is organized as follows. 
In Section 2, we give a detailed description of our physical model. 
We describe the DM simulation parameters, and how we generate DM halo catalogs in Section 3. 
In Section 4, we first construct \lya\ LF using model LAEs at $z\approx$  3 and compare it to the observations to
 find the best-fit model parameter, and then use this best-fit parameter to construct \lya\  LFs at $z >$ 3. 
 In Section 5, we derive the physical properties of LAEs using our best-fit model, estimate the dust mass in our model 
 LAEs to compare with the dust estimates from observations, construct UV LF of our model LAEs 
 and compare it with the observations, and then investigate the evolution of \lya\ LF. 
 We study the large-scale structure of model LAEs, and study the redshift evolution of correlation 
 lengths of LAEs in Section 6. We summarize and present conclusions in Section 7.

\section{Physical Model for \lya\ Emitters}

Our model is motivated by the idea that \lya\  emission in LAEs is associated with star formation 
\citep{pa67} 
 from rapid accretion of new material on to the DM halos. 
This new material provides fresh fuel to the system driving the star formation  \citep{ker09}.

We populate each DM halo with an LAE, and assign to it \lya\  luminosity ($L_{Ly\alpha}$) proportional to the 
SFR using the following equation:
\begin{equation}
L_{Ly\alpha}= 1 \times 10^{42} \times \frac{SFR}{M_{\odot} {\rm yr^{-1}}} \; \; {\rm erg} \; {\rm s}^{-1}.
\end{equation}
Here $L_{Ly\alpha}$ is the intrinsic luminosity of an LAE. The observed  \lya\ flux will also depend on the 
escape fraction of the \lya\ photons  ($f_{esc}^{Ly\alpha})$. 
Moreover, equation (1) implicitly assumes an escape
 fraction near zero for the ionizing continuum photons, whose absorption is required to 
 produce the  \lya\ emission line.

While the escape fraction of ionizing Lyman continuum photons ($\lambda < 912 \; \rm{\AA}$) 
is not very precisely known, 
several studies have shown that an escape fraction of only a few percent is sufficient to meet the 
observational constraints on the reionization epoch 
 \citep[e.g.][]{woo00,ha06,raz06,gne08}.
In addition, observations are also generally consistent with small escape fractions of Lyman continuum 
photons both locally 
\citep{le95} and at high redshifts \citep[e.g.][]{ ha06, sha06}.

The escape fraction of \lya\ photons ($f_{esc}^{Ly\alpha}$) is likely to be large with $f_{esc}^{Ly\alpha} \approx 1$
 causing the large 
observed \lya\ EWs 
\citep{mr02, fi07, wan09}.
 However, 
the semi-analytic model of 
\citet{le06}
 predicts a much smaller value of $f_{esc}^{Ly\alpha} =2\%$  
which is compensated by top-heavy initial mass function in their model. For simplicity, in our model 
we approximate 
 $f_{esc}^{Ly\alpha}$ and Lyman continuum photons as unity and zero, respectively. 
 Thus, all the \lya\ 
photons produced in LAEs escape to be observed while none of the ionizing photons escape from the galaxy. 
Small deviations from these assumptions will not affect our results significantly.

As stated earlier, in our model we assume that the accretion of new material on to DM halos causes 
star formation in LAEs. We estimate the SFR, \textit{i.e.} the mass in accreted gas $\Delta M_{gas}$ 
converted to stars in unit time, in LAEs by converting baryonic mass accreted $(\Delta M_{b})$ by DM halos, 
adopting a constant ratio of baryons to the DM, over a short timescale, $t_{Ly\alpha}$. This timescale  ($\approx 30$ Myr )
 is broadly similar to the stellar population ages of most \lya\  galaxies 
\citep[e.g.][]{pi07, fi07, fi08},
 the lifetimes of OB associations, and the dynamical time expected for \lya\ galaxies based on their measured
  sizes
   \citep{pi07}.
 In addition, the dust produced in supernovae (SNe), which occurs approximately on timescales 
 of  $\sim$ 30 Myr, may reduce the fraction of \lya\  photons escaping from LAEs, thus giving rise to the 
 duty cycle of LAEs
  \citep{ko07},
   which is also reproduced in our model. A similar timescale 
   ($\approx$ 70 Myr)  was used by 
    \citet{shi07}
    to match the morphology of large-scale structure of LAEs by varying the amplitude of density 
    fluctuations in galaxy formation models.

Thus,
\begin{equation}
SFR= f_\star \times \left( \frac{\Delta M_{gas}}{t_{Ly\alpha}}\right)=  
f_{\star} \times \left( \frac{\Delta M_{b}}{t_{Ly\alpha}} \right)= 
f_\star \times \dot{M_b}~~,
\end{equation}
where $f_\star$  is the SFE. In the above equation, we have assumed that $\Delta M_{gas}$ is same 
as the baryonic mass accreted ($\Delta M_{b}$) by the DM halos, and since our simulation contained 
only DM particles, we use the universal ratio of baryonic and DM densities \textit{i.e.,}
$\Delta M_b= ({\Omega_b}/{\Omega_{\rm DM}}) \times \Delta M_{\rm DM} $,
where  $\Omega_b$ and $\Omega_{\rm DM}$ are the baryonic and DM 
density parameters and $\Delta M_{DM}$ is the dark matter mass accreted by the DM halos.

Finally, the total mass in young stars in a LAE, is estimated using
\begin{equation}
M_{\star} \approx SFR \times t_{Ly\alpha} =
f_{\star} \times \dot{M_b} \times t_{Ly\alpha} =
 f_{\star} \times  \frac{\Omega_b}{\Omega_{\rm DM}} \times \Delta M_{\rm DM} ~~.
\end{equation}
This corresponds to the mass of stars younger than 30 Myr which contribute to the \lya\ and UV continuum 
emission, which is more easily measured than the total stellar mass. 
The only unknown variable 
in all of the above equations is $f_\star$,
the only free parameter in our model. 
Here we note that 
$f_{\star}$ and $\Omega_b/{\Omega_{\rm DM}}$
 ratio in the above equations are degenerate and these values may vary somewhat for individual galaxies.

\section{Simulation \& Halo catalogs}

We constructed the DM halo catalog using an N-body 
DM cosmological simulation code GADGET2 \citep{spr05}. 
We generated the initial conditions for the simulation using 
 second-order Lagrangian Perturbation Theory \citep{cro06,tha06}.
In this simulation we use $1024^3$ DM particles in a comoving volume of (102  Mpc)$^{3}$, 
a volume greater than a typical LAE survey.
Each DM particle has a mass $\approx  2.7\times  10^{7} M_{\odot} h^{-1}$. 
Using a Friends-of-Friends (FOF) halo finder \citep{dav85}, we identify DM halos that contain 100 or more DM
particles. 
This corresponds to a minimum halo mass  $\approx  2.7\times  10^{9} M_{\odot} h^{-1}$.
We then generate  catalogs, for redshifts from  $z=10$ to $z=3$,  which contain positions of 
halos, their  DM mass, and unique IDs of each individual particle that belongs to a given halo.
These unique particle IDs are later used to track halos between two epochs.
Throughout this work we assumed a flat $\Lambda$CDM cosmology with parameters 
$\Omega_{m}$=0.233, $\Omega_{\Lambda}$=0.721, $\Omega_{b}$=0.0462,
$h$=0.71, $\sigma_{8}$=0.817 where $\Omega_{m}$, $\Omega_{\Lambda}$,
$\Omega_{b}$, $h$,  and $\sigma_{8}$ correspond, respectively, to the matter density,
dark energy density, and baryonic density in units of the critical density,
the Hubble parameter in units of 100 km s$^{-1}$ Mpc$^{-1}$, and the RMS density
fluctuations on the 8 Mpc $h^{-1}$ scale, in agreement with WMAP \citep{spe07} five year
results \citep{hin09}.

\begin{figure}[t!]
\epsscale{1.2}
\plotone{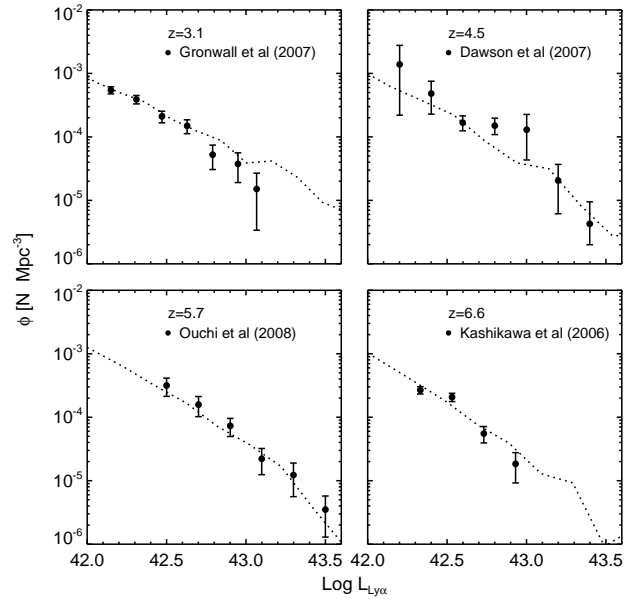}
\caption{ Evolution of  \lya\ LFs at redshifts $z\approx 3-7$. The dotted lines show  results from our model
and the symbols with error bars are the observational  data. (a) The best-fit model  \lya\ LF at $z=3.1$
yields a SFE of 2.5$\%$. We use this SFE  to construct model
\lya\ LFs at $z$=4.5, 5.7, and 6.6 (b)-(d).
The references for the data: $z=3.1$ 
\citep{gro07}, $z=4.5$ \citep{da07}, $z=5.7$ \citep{ouc08}, and $z=6.6$ \citep{ka06}.
}
\end{figure}

\section{Lyman Alpha  Luminosity Function} 
We now construct the \lya\ LF, the number of LAEs per unit volume in a given luminosity bin. 
First, we calculate the total DM mass accreted ($\Delta M_{DM}$) by each DM halo at  $z=3.1$ during an 
interval $\approx$30 Myr (equals $t_{Ly_{\alpha}} $ in equation (2)). 
To calculate $\Delta M_{DM}$ we track each halo, using the unique ID associated with particles in a given halo, 
between two epochs separated in time $t_{Ly_{\alpha}} $. In general, we expect every halo to accrete
 more mass with time. However, due to group finding noise, we find that some halos lose mass (negative mass accretion) between outputs. In other words, the mass accretion by some halos is not real but results from the simulation noise. The main reason for this noise is the way halos are identified in any DM simulation.

\begin{figure*}[t!]
\epsscale{0.7}
\plotone{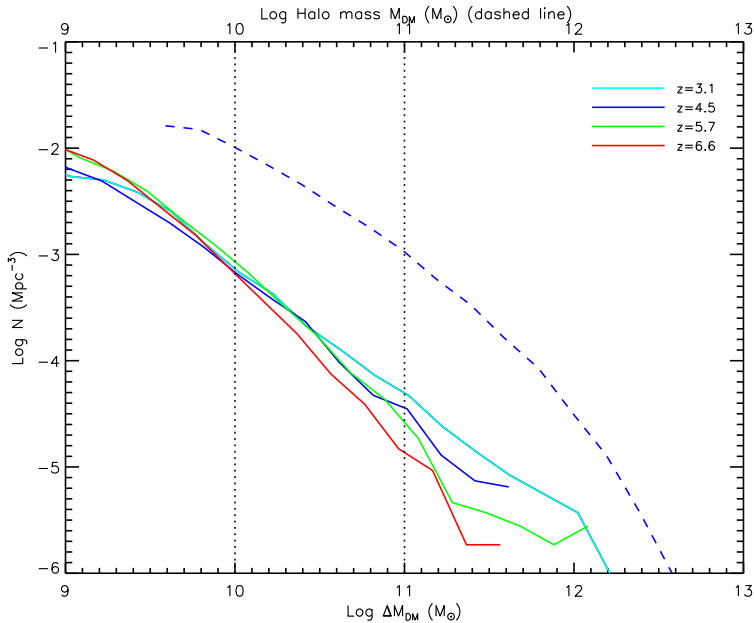}
\caption{ Accreted mass function and halo mass functions. Solid lines show accreted mass
functions at $z=3.1$ (violet), $z=4.5$(blue), $z=5.7$ (green) and $z=6.6$(red). The dashed 
blue line shows the dark matter halo mass functions at $z=4.5$ to compare with the corresponding
accreted mass function.  The vertical dotted lines enclose the region of observed \lya\ luminosities 
in LAEs.}
\end{figure*}

In our DM simulation, we use a FOF halo finder which links all the particles within a linking length from each other into a halo, independent of whether a given particle is gravitationally bound to a halo. 
Thus, associating a particle with a halo based on the linking length gives rise to some uncertainty in 
halo mass (in this case $\Delta M_{DM}$). To determine how many halos have real accretion, rather than 
spurious apparent accretion due to uncertainty in particle association with a halo by the halo finder, 
we first construct a histogram of $\Delta M_{DM}$  including the halos with negative $\Delta M_{DM}$. 
We then subtract the halo counts in negative $\Delta M_{DM}$ bins from the corresponding counts in the positive 
 $\Delta M_{DM}$bins. This procedure compensates for halos that show accretion just due to random nature of a FOF halo finder. The remaining halos with positive accretion rates are then considered for constructing \lya\ LFs.

Next, we convert the accreted mass bins to the \lya\ luminosity bins using equation (1) to yield 
\lya\ LF. We then compare this LF with the observations at 
$z=3.1$ \citep{gro07},   and get the best-fit model by varying the SFE ($f_\star$)  to yield the least reduced 
$\chi_{r}^{2}$ ($\chi^{2}$ per degree of freedom) given by
\begin{equation}
\chi_{r}^{2}=\frac{1}{N-1}\sum \frac{(N_{model} -N_{obs})^{2}}{\sigma_{model}^{2}+\sigma_{obs}^{2}},
\end{equation}
where $N$, $N_{model}$ \& $N_{obs}$  are the number of observed data points, 
LAE counts from model, and the observed LAE counts in each bin, respectively, and the 
Poisson errors are given by  
$\sigma_{model}=\sqrt{N_{model}}$ and  $\sigma_{obs}=\sqrt{N_{obs}}$.
Figure 1 (top left) shows the best-fit model \lya\  LF (dotted line) at z = 3.1. 
The symbols are the observations from \citet{gro07} shown with 1$\sigma$ error bars.

Lastly, we use the best-fit model parameter  $f_{\star}$ \textit{i.e.,}
 the SFE at z = 3.1 to construct the model \lya\  LFs at $z>3$, 
 and then compare these LFs with the observations at $z=$ 4.5, 5.7, and 6.6. 
 Figure 1 shows the \lya\  LFs from our model (dotted lines) and observations (filled circles) at redshifts 
 $z=3.1$ \citep{gro07}, $z=4.5$ \citep{da07}, $z=5.7$ \citep{ouc08}, and $z=6.6$ \citep{ka06}. 
 We have rebinned the observational data for 
  $z=3.1 $ and $z$=6.6 
 data so as to make the bin size uniform at all redshifts. 
 Wiggles seen, especially in the 
 $z=4.5$
 model \lya\  LF (Figure 1, top right) are probably due to statistical noise. 
 The best-fit model for $z=3.1$  \lya\  LF yields a SFE of 2.5$\%$. 
 Corresponding to this SFE, the 
 $\chi_{r}^{2}$ values between our model and 
the observed \lya\ LFs are 0.5, 0.8, 1.2 $\&$ 1.5 for \lya\ LFs at $z=$3.1, 4.5, 5.7, and 6.6, respectively.

\begin{figure*}[t!]
\epsscale{0.7}
\plotone{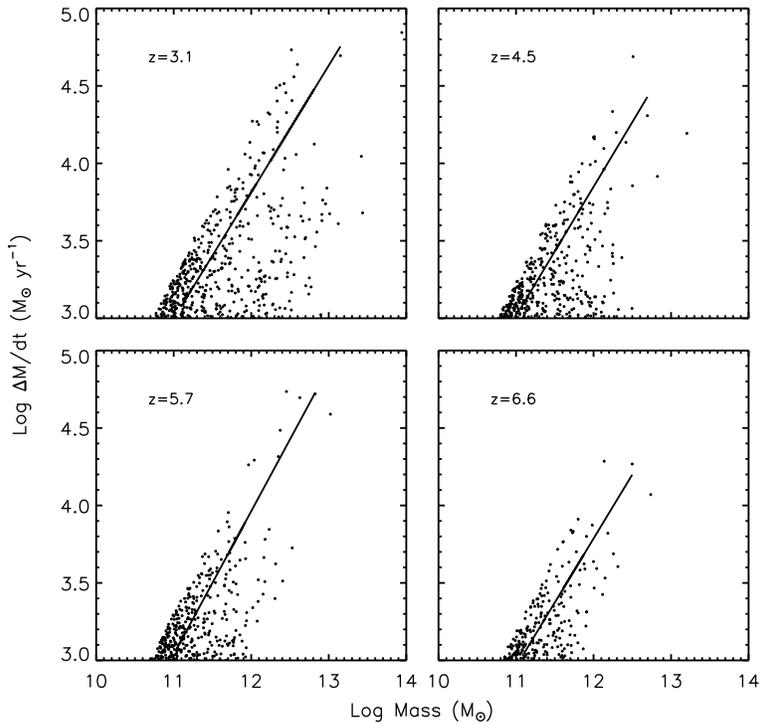}
\caption{ The DM accretion rate of halos as a function of halo mass at $z\approx 3-7$.
The solid line is the least-square fit to the median mass in each $\Delta M/dt$ bin.
The slope of the lines is nearly constant $\approx 0.8-0.9$ over all redshifts.}
\end{figure*}

\section{Results}

Our model  \lya\  LFs with single SFE, agree remarkably well with the observations  \citep{da07,ouc08}. 
They reproduce, without any additional parameters, the duty cycle of  $\sim$ 10$\%$
 obtained from clustering studies \citep{ko07}.  
To predict the  \lya\ LFs of LAEs,  \citet{na08} 
 investigated two models, the duty cycle and escape fraction scenario, and found that the duty cycle 
 model reproduces observations better than the escape fraction model. In our model, the duty cycle is 
 naturally produced since only halos with high accretion rates will be observed as LAEs. Figure 2 shows the 
 halo mass function at z = 4.5 (blue dashed line) and the accreted mass function (blue solid line). 
 \textit{Thus, the use of accreted mass rather than the total halo mass eliminates the need to introduce an additional duty cycle parameter in our model.}

LAE observations at high redshifts suggest that many of the properties of LAEs such as the LFs do not evolve 
significantly over a wide redshift range  \citep{da07, ouc08}.
These observations are in agreement with our model predictions,\textit{i.e.}, our model predicts nearly a constant SFE 
over a wide redshift range and that other physical properties including \lya\  luminosity, and SFRs do not evolve 
significantly from $z\approx$ 3 to 7 since, in our model, these properties depend on SFE. \citet{jim05}
 also found a similar constant SFE over a wide range ( $\approx$ 2 orders of magnitude) of stellar masses, 
 and over a relatively large redshift range, using a large SDSS spectroscopic sample of galaxies 
 at $z< 0.3$, combined with stellar population models. 
 Figure 3 shows the DM accretion rate as a function of halo mass. 
 The solid line is the least-square fit to the median mass in a $\Delta M/dt$ bin. 
 A nearly constant (0.8-0.9) slope of this line at all redshifts implies that the SFR does not evolve in this 
 redshift range, while deviation of the slopes from unity suggests that the mass accretion rate is a nonlinear 
 function of halo mass. Using least-square fits to the median mass in each 
$\Delta M/dt$ bin (Figure 3), we obtain an average DM mass accretion rate  $\Delta M/dt$,
\begin{equation}
\frac{\Delta M}{dt} \approx 4.3 \times 10^{-7} M_{\rm DM}^{0.85} \;  M_{\odot}  \rm{yr^{-1}},
\end{equation}
where $M_{\rm DM}$  is the DM halo mass. 
For example, for a halo mass of $10^{11}$ M$_{\odot}$, the baryonic mass accretion rate, obtained by 
converting DM mass accretion rate using universal ratio of baryons to DM, is approximately  
170 $\;  M_{\odot}  \rm{yr^{-1}}$.

\subsection{Physical properties of LAEs}
We use the best-fit model parameter (SFE) at $z=3.1$ to derive other physical properties of LAEs and
 compare our results with the observations. Our best-fit model yields an SFE of 2.5$\%$, 
 consistent with the global SFE. 
 \citet{fuk98}
 predicted a SFE $<5\%$, while  \citet{bal08} 
 found this value in the range  $4\%-8\%$ for blue light in galaxies. 
 While our model predicts a roughly constant SFE over a wide redshift range, 
 in reality this value will vary somewhat depending on the ratio of baryons to DM.

The SFE of 2.5$\%$ yields SFRs  $\approx 1-10 \; M_{\odot}$ yr$^{-1}$ 
 corresponding to the observed \lya\ luminosity range  $L_{Ly\alpha} \approx 1 \times 10^{42} - 1\times 10^{43}$ erg s$^{-1}$.
  This SFR is comparable to the inferred SFR $\approx  8\; M_{\odot}$ yr$^{-1}$ in 
LAEs at $z\approx 5$  \citep{pi07}.
A similar average value of SFR $\approx  6\; M_{\odot}$ yr$^{-1}$  was inferred for $z=3.1$ LAEs 
\citep{ga06}.   A slightly higher value of SFR $ \approx  5.7 - 28.3 \; M_{\odot} $yr$^{-1}  $
 with median SFR  $ \approx  9.6 \; M_{\odot} $yr$^{-1}  $  was inferred for $z=5.7$
 LAEs \citep{mur07}. For $z=6.6$ LAEs,  \citet {ta05}  found an average SFR $ \approx  5.7 \pm 2.3\; M_{\odot} $yr$^{-1}  $.
 These averages however depend on the depth of the surveys;  
  deeper surveys probe less luminous galaxies and hence lower SFRs.
The total stellar mass in young stars (estimated using Equation (3)) of LAEs corresponding to the observed 
\lya\  luminosity range is $M_{\star} \approx (3 \times 10^{7} )- (3 \times 10^{8})M_{\odot}$ 
 in good agreement with the observed stellar masses    $\approx$  $10^{7} M_{\odot} -
10^{9} M_{\odot}$ of LAEs  \citep{fi07, ga07,  pi07, pen09}.
Thus, our model reproduces the primary physical properties of LAEs at  $z \approx 3-7$.
In addition, our assumption of large escape fraction of \lya\  photons will yield high observed EWs of \lya\ 
line in LAEs \citep{mr02, kud00, da04, da07, shi06, wan09}.

\begin{figure*}[t!]
\epsscale{0.8}
\plotone{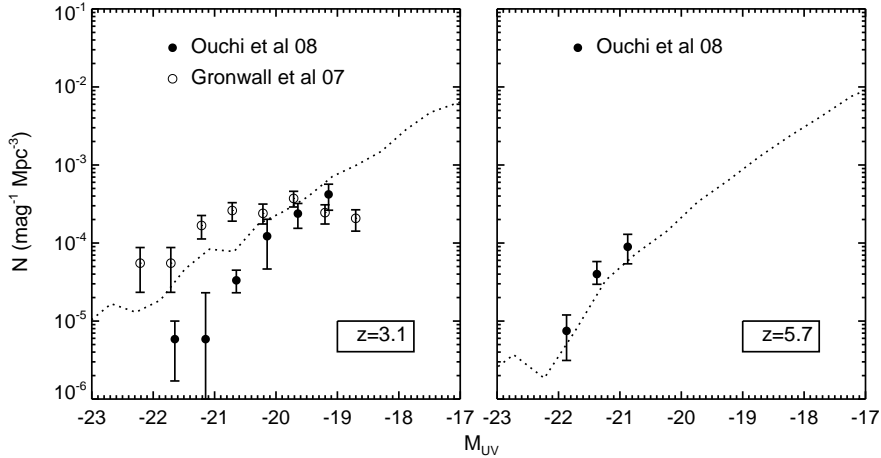}
\caption{ 
UV LFs of LAEs at $z = 3.1$ and $z = 5.7$. The filled and open circles are the data from \citet{ouc08}
\& \citet{gro07}, respectively. The dotted line is our model predicted UV LF of LAEs.}
\end{figure*}

\subsection{Dust mass in our model LAEs}
In our model we  assume that   LAEs do not contain significant amount of dust;  however,
we can estimate   dust masses for our model LAEs using the dust estimates  from SNe
\citep{bia07}.
Assuming a SNe rate $\approx 1/ 150 \; M_{\odot} $ \citep{sca05} and that each SN produces 
$\sim$ 0.1-0.6 $M_{\odot} $ of dust \citep{bia07} of which nearly $2\%-20\%$ survives  \citep{bia07},
we estimate a dust mass  $M_{\rm dust}$ $\approx (4\times 10^{3})-(2 \times 10^{5})\; M_{\odot} $
for our model LAEs with SFR $\approx$ 10 $M_{\odot}  \rm{yr^{-1}}$.

We now compare these values with the dust mass of LAEs inferred from observations.
With extinction of $A_{V}$=0.1-1.5 for LAEs at $z\approx 4.5$ \citep{fi08} , with their sizes of 1kpc in radii
\citep[S. Malhotra et al 09 in preparation]{bon09}, and assuming
a dust to gas ratio of 1/200, we estimate a dust mass $\approx (3 \times 10^{4}) -(4.5 \times 10^{5}) \; M_{\odot}$, 
in agreement with the estimated values for our model LAEs.
There are, however considerable uncertainties in both, observational and theoretical estimates of dust mass.

\subsection{UV Luminosity Function of LAEs}
Our model, with single parameter, \textit{i.e.}, the SFE, reproduced the observed \lya\  LFs over a wide range of 
redshifts, $z = 3-7$. Now we compare the UV LFs of our model LAEs with the observations at $z$ = 3.1 
\citep{gro07, ouc08} and $z$ = 5.7 \citep{ouc08}.
We convert \lya\ luminosity to UV luminosity using the following relation \citep{mad98}
\begin{equation}
L_{UV} (erg s^{-1} Hz^{-1})=8 \times 10^{27} \times SFR (\;  M_{\odot}  \rm{yr^{-1}}),
\end{equation}
where $L_{UV}$ is the UV luminosity at $\rm 1500 \AA$, and SFR is the star formation rate of our model 
LAEs calculated using Equation (2). 
Figure 4 shows the comparison between model predicted and observed UV LFs at $z$ = 3.1 and $z$ = 5.7. 
Filled and open circles are the observations from \citet{ouc08} and \citet{gro07},
 respectively, while the dotted line is our model predicted UV LF. At $z$ = 5.7, the model predicted UV LF 
 agrees quite well with the observations, while at $z$ = 3.1, the observed UV LFs of \citet{ouc08}  and 
 \citet{gro07}  brackets our model predicted UV LF.

\subsection{Evolution of \lya\ luminosity function}

The \lya\ LFs have  been used to probe the epoch of reionization 
and constrain the evolution of intergalactic medium (IGM) 
\citep[e.g.][]{ha99, mr04, hai05,st05, ka06,di07, mcq07,mes08, ota08}.
Any significant evolution in the number density of LAEs, after accounting the newly formed LAEs 
between two  redshifts will imply that the IGM evolved at these redshifts.

Currently, the evolution of \lya\ LF at $z>5$ is not well understood.
Previous studies   find no significant evolution 
in  \lya\ LFs at redshifts  between  $z=5.7$ and $z=6.5$ \citep{mr04}. 
However recent observations  \citep{ka06, ota08}  find a modest   decline in the bright end of 
the \lya\ LF  from $z=5.7$ to $z=6.6$ suggesting IGM evolution at these redshifts.
\citet{di07} showed that the weak \lya\  LF  evolution between $z=5.7$ \& $z=6.6$ can be 
attributed to the evolution of DM halo mass function.
In addition, the cosmic variance in a volume limited LAE survey also affects the \lya\
LF. 
For example, \citet{shio09} find that the number density of LAEs, at $z\approx5$  vary by 
a  factor  $\approx 2$ in a survey area of $ 0^{\circ}.5 \times 0^{\circ}.5$. 
We now  investigate the evolution of \lya\ LFs, and the effect of cosmic variance on number 
density of LAEs in a volume and flux limited LAE survey.

 \begin{figure}
\plotone{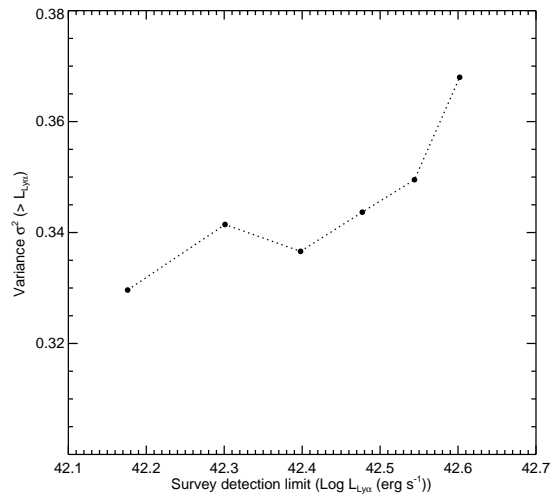}
\caption{ Field-to-field variance of number of LAEs at $z=6.6$, measured in eight subvolumes 
(102$\times$51$\times$25 Mpc$^{3}$), plotted as a function of survey detection limit. 
The variance $\sigma^{2}>30\%$ for a typical narrowband LAE survey with  \lya\ detection limit 
$>2\times10^{42} $ erg s$^{-1}$.}
\end{figure}

From Fig.\ 2 we see that there is some  evolution of number density of DM halos that can host 
LAEs. However, this
weak evolution is due to the intrinsic change in the number density since we have not  included
any IGM correction in our model.
 Thus our model can explain the observed evolution of \lya\ LF \citep{ka06} without invoking
sample variance or reionization.
We now estimate the  effect of sample  variance on  the \lya\ LF.

\begin{figure*}[t!]
\epsscale{0.9}
\plotone{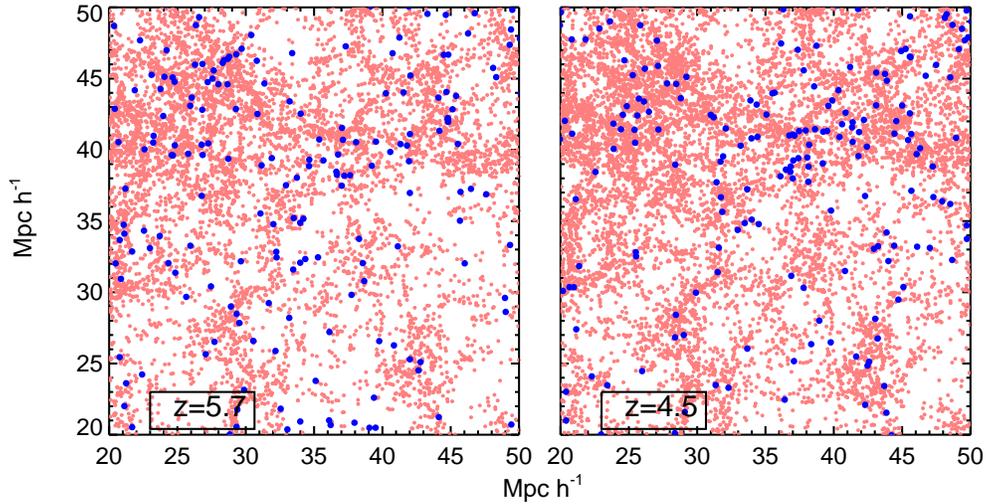}
\caption{ The spatial distribution of our model LAEs in a slice from DM simulation at two 
redshifts   $z=5.7$ (left) and $z=4.5$ (right) in a volume $30 \times 30 \times 17 \; h^{-3}$  Mpc$^{3}$. 
 The small (red)  and big (blue) filed circles represent  the positions of  DM halos and model LAEs, 
 respectively. 
  Only LAEs with $ L_{Ly\alpha}> 2\times10^{42} $ erg s$^{-1}$ are plotted. 
  In general, the LAEs are located in  high  density regions.  Also  note that different halos
  host  LAEs at the two redshifts, depending on whether they are accreting or not. This gives rise to 
  a duty cycle quite naturally.   }
\end{figure*}

In order to  understand  the effect of cosmic variance  on the observed number density  
of LAEs in a volume and flux limited LAE survey, we estimate the field-to-field variance 
by dividing the total simulation volume into eight non-overlapping rectangular boxes, each with a 
comoving volume (102$\times$51$\times$25 Mpc$^{3}$) comparable to typical narrow-band LAE 
surveys $(\approx 2\times 10^{5} \; \rm Mpc^{3})$.
The variance is calculated using $\sigma^{2}=<(N-\mu)^{2}>/\mu$, where N is the 
number of LAEs in each sub-volume and $\mu$ is the average number of LAEs. 
Figure  5  shows  field-to-field variance, for a volume limited LAE survey with different \lya\ flux limit. 
For a narrow-band LAE survey with  \lya\ detection limit of $ L_{Ly\alpha}> 2\times10^{42} $ erg s$^{-1}$
and a survey volume $ < 2 \times 10^{5} \; \rm{Mpc^{3}}$,  the field-to-field variation is significant with 
 $\sigma^{2} \gtrsim 30\%$.
This result confirms the necessity of using a large volume to minimize sample variance in LAE
surveys.
It also   strengthens our conclusion that some apparent evolution of \lya\ LF can be attributed to 
the sample variance.

\section{Clustering of LAEs}

Previous studies \citep[e.g.][]{ili08, ors08}  suggest that  LAEs  trace rarer and higher density regions.
However, \citet{shi07}  suggested that the LAEs at $z\approx 3$ do not 
necessarily reside in high density peaks.
In this section, we investigate whether our model LAEs reside in high density regions and 
 estimate their spatial correlation lengths.
Figure  6 shows the spatial distribution of LAEs at two different redshifts, $z=5.7$ and $z=4.5$, 
in a simulation  slice of 30$\times$30$\times$17 $h^{-3}$ Mpc$^{3}$.
The depth of this slice is comparable to the depth  of a typical LAE survey.
Comparing the locations of LAEs at two redshifts it is clear that  
different halos host LAEs at different redshifts, thus exhibiting a duty cycle \citep{ko07,st07, na08}.
Our model LAEs are generally located around overdense regions consistent with the 
observations and as expected in  biased galaxy formation models.

\subsection{Two-point spatial correlation function}

The two-point spatial correlation function $\xi(r)$ \citep{peb80}
is frequently used to study the clustering properties
of galaxies \citep[e.g.,][]{ze05, ga07,ko07}. 
We use the Landy-Szalay estimator, proposed by \citet{la93},
to calculate the two-point 
spatial correlation function given by
\begin{equation}
\xi(r)=\frac{DD(r) -2DR(r) + RR(r)}{RR(r)},
\end{equation}
where $DD(r)$, $RR(r)$, and $DR(r)$ are the number of galaxy-galaxy, 
random-random, and galaxy-random pairs, respectively with separation distance of  $(r, r+\delta r)$.
To compare our model  $\xi(r)$ with the observations and quantify its  evolution with redshift  
we only include LAEs brighter than the detection limit of LAE surveys at a given redshift, 
and use  our full simulation volume $\approx 1\times 10^{6} \; \rm{Mpc^{3}} $ for better statistical
significance.

To calculate   $\xi(r)$ at each redshift, we generated a random sample of points
with uniform  coordinates drawn from a uniform probability
distribution, and a number of random points exactly equal to the
number of LAEs.  
We count the number of pairs, DD(r), RR(r), and DR(r)
separated by a distance $r$ by binning the points at
different $r$ with binwidth of $\delta r=0.2h^{-1}$ Mpc.  
To minimize the random errors, we perform 50 realizations with different sets of random points
 and calculate an average $\xi(r)$ at  each $r$. 
Using  $r< 20 \; h^{-1}$ Mpc and assuming negligible error, we obtain  the spatial correlation length $r_{0}$  by fitting a least-square power law  to the correlation function.

\begin{figure*}
\epsscale{0.9}
\plotone{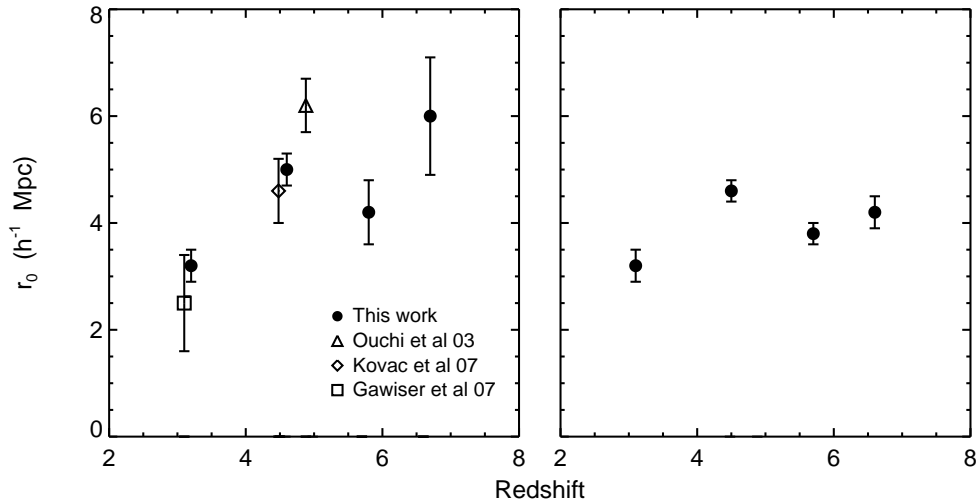}
\caption{ Correlation lengths of LAEs at different redshifts.
\textit{Left}.
Comparison of correlation lengths of our model and observed LAEs at different 
redshifts.  Here we include LAEs with \lya\ luminosity greater than the survey limit at each 
redshift. The filled circles are our model results, other symbols are from different 
observations as shown in the labels. 
The observed $r_{0}$ values shown here for $z$=4.5 \& 4.86 are for the contamination corrected (the maximum
value permitted) LAE sample.
Our model results are slightly shifted to avoid overlap 
with  other observational  points.
\textit{Right}: The correlation lengths of our model LAEs with a constant \lya\ luminosity cutoff 
($ L_{Ly\alpha}> 1\times10^{42} $ erg s$^{-1}$) at all redshifts, showing that the apparent evolution of correlation
length with redshift (seen in left panel) is mostly due to different luminosity detection limits in LAE surveys.
}
\end{figure*}

The correlation lengths obtained from our model LAEs  show a modest evolution, with 
$r_{0}=( 3.2 \pm 0.3, 5.0 \pm 0.3 ,  4.2 \pm 0.6 , 
6.0 \pm 1 )h^{-1}$ at  $z=$(3.1, 4.5, 5.7,  6.6), respectively.
Figure 7 (left) shows a comparison between our model predictions  (filled circles) and observed 
$r_{0}$ (shown with different symbols given in the labels) at different redshifts.
The model predicted $r_{0}$ values are slightly  shifted along x-axis to avoid overlap with the observations.
The  predicted correlation lengths are  consistent with the observations at  $z=3.1$ with observed 
$r_{0}=$ 2.6 $\pm$1 $h^{-1}$ 
\citep{ga07}, and at $z=4.5$ with observed  $r_{0}=$ 4.6 $\pm$0.6 $h^{-1}$ \citep{ko07} estimated using
contamination-corrected  (the maximum value permitted) LAE sample.
 However, \citet{ouc03} found a higher  $r_{0}=6.2\pm0.5 $ $h^{-1}$ Mpc for contamination-corrected  LAE 
 sample $z=4.86$.
We now  estimate the variance  of  $r_{0}$ in a volume and flux limited LAE survey, and see if 
we can account for the  large difference seen in the observed $r_{0}$  between  $z=$4.5  \& $z=$4.86 
LAE surveys.

To estimate the variance of $r_{0}$  ($\sigma^{2}_{r_{0}}$) in a volume and flux limited LAE survey, 
we divide the total simulation
volume at $z=$3.1 into five non-overlapping sub-volumes, each with a comoving volume of 
(102$\times$102$\times$
20.4  Mpc$^{3}$), approximately equal to a typical survey volume and only including LAEs with 
$ L_{Ly\alpha}> 1\times10^{42} $ erg s$^{-1}$.
We calculate $\xi(r)$ and $r_{0}$,  as described above (second paragraph) in each sub-volume and estimate
$\sigma^{2}_{r_{0}}$  at $z=$3.1.
We find  $\sigma^{2}_{r_{0}}$=0.5 $h^{-1}$ Mpc with an average $r_{0}$=3.2 $h^{-1}$ Mpc, average of  $r_{0}$
in five sub-volumes.
While this variance in $r_{0}$  cannot account for the large difference
in   $r_{0}$  observed between the  two surveys at $z=$4.5 \& 4.86, it is clear that one needs to take into account such variance
in correlation lengths obtained from  volume and flux limited  surveys.

Finally we investigate if the redshift evolution of $r_{0}$ seen in Figure  7 \ ({\textit{left} panel) is significant since this
 can result from the   surveys at lower redshifts extending to lower luminosities and  hence 
probing lower halo masses.
In order to understand this effect, we consider full simulation volume and only include LAEs with a constant 
$ L_{Ly\alpha}> 1\times10^{42} $ erg s$^{-1}$ at all redshifts. Choosing a constant   luminosity cutoff at 
all redshifts  implies that we are probing approximately same halo masses at all redshifts.
We calculate $r_{0}$ in the same way as described above (second paragraph) except that we impose  the same 
luminosity cutoff at all redshifts.
Figure  7 \ (\textit{right} panel) shows $r_{0}$ with a constant lower luminosity at all redshifts.
Most of this evolution seen in Figure  7 (\textit{left} panel) can be attributed
to the luminosity limit of different surveys probing different halo masses.

\section{Summary and Conclusions}
We have used a physical model with a single variable parameter to populate DM halos with LAEs in a 
cosmological simulation and compared our model predictions with the observations at redshifts $z\approx 3-7$. 
In our model, we assumed that the SFR, and hence the \lya\  line luminosity is proportional to the mass accreted by halos. 
In other words, the star formation in LAEs mainly results from the accretion of new material. 
Despite the lack of observational evidence relating accretion rate to the \lya\  luminosity of LAEs, 
it is promising that our model gives a good fit to the observations over a wide range of redshifts and is able to 
reproduce several other physical properties of LAEs. In addition, relating the accreted mass rather than the total halo 
mass to the \lya\ luminosity gives rise to a duty cycle of LAEs quite naturally.

To compare our model predictions with observations, we first constructed the  \lya\ LF at $z=3.1$ and obtained the best-fit model by varying the SFE and comparing the model  \lya\ LF with the observations at this redshift. We then used this best-fit model to predict the  \lya\ LFs and physical properties of LAEs at $z=4.5, 5.7$ and $z=6.6$.

Using a constant SFE, our model predicted \lya\ LFs agree remarkably well with the LAE observations over a wide
 redshift range. Our best-fit model yields a SFE = 2.5$\%$ which gives SFR $\approx 1- 10 \;M_{\odot} $yr$^{-1}$
  in good agreement with the observations. 
  We find that the model LAEs in the currently observable luminosity range 
  ($2 \times 10^{42} \la L_{Ly \alpha}  \la 2 \times 10^{43} {\rm erg} \;  {\rm s}^{-1}$)
  have stellar masses $\approx 3 \times 10^{7}$ to  $3 \times 10^{8} M_{\odot}$
  of young ($<30$ Myr) stars. 
  These stellar masses of LAEs are similar to those inferred from observations  \citep{fi07, pi07}.
We have estimated the dust mass 
$M_{\rm dust}$ $\approx (4\times 10^{3})-(2 \times 10^{5})\; M_{\odot} $
 for our model LAEs with $L_{Ly\alpha} = 1 \times 10^{43} $ erg s$^{-1}$,
in agreement with the inferred dust masses $\approx (3 \times 10^{4}) - (4.5 \times 10^{5}) \; M_{\odot}$
from LAE observations at $z\approx4.5$ \citep{fi08}.
Using our model LAEs, we constructed UV LF and compared it with the observations 
at$ z = 3.1$ and $z = 5.7$. At $z = 5.7$, our model predicted UV LF of LAEs agrees quite well with the observations, 
while at $z = 3.1$, the observed UV LFs of \citet{gro07} and \citet{ouc08}
bracket our model-predicted UV LF.

While our model predicts a constant SFE, and hence a weak evolution of other physical 
properties of LAEs  over a redshift range $z=$3-7, this value also depends on the ratio
of baryons to DM. 
Thus, in reality $f_{\star}$, and hence other physical properties of LAEs might show detectable, albeit
weak, evolution with redshift.

We also investigated the evolution of \lya\ LFs from $z\approx 3-7$ and find that 
there is no significant evolution of \lya\ LF due to the IGM if we  include the intrinsic change in the 
number density of LAEs over this redshift range.
This conclusion is strengthened  if we include the effect of cosmic variance on the observed
number density of LAEs.
We show that the field-to-field variance can be large $\approx 30\%$  for a flux and volume
limited surveys comparable to current observations.

We studied the clustering properties of LAEs and  found that the LAEs are mostly located in the 
high density peaks.
Our model predicted correlation lengths  $r_{0}=( 3.2 \pm 0.3, 5.0 \pm 0.3) h^{-1}$ Mpc) which are  in good 
agreement with the observations.
We also estimate the variance ($\sigma^{2}_{r_{0}}$)  in $r_{0}$ for a volume and flux limited LAE survey
and find that $\sigma^{2}_{r_{0}}$=0.5 $h^{-1}$ Mpc at $z=$3.1.
Our models predict a modest evolution of the correlation length with redshift.
Currently, there are no measurements of   $r_{0}$ at $z>5$ due to insufficient sample size of LAEs 
at higher redshifts.
Therefore,  more  data are needed to test our predictions at higher redshifts in order to understand the  
evolution of  $r_{0}$ with redshifts.

\acknowledgments
We acknowledge helpful suggestions from Kyoung-Soo Lee. We are also grateful to the
anonymous referee for insightful comments and suggestions.
This work was  supported in part by the School of Earth \& Space Exploration, 
Arizona State University, the NSF grant AST-0808165,  the Swiss National Science
Foundation grant 200021-116696/1, and Swedish Research Council
grant 60336701.
R. J. T. is supported by a Discovery Grant from NSERC, the Canada Research Chairs program and
the Canada Foundation for Innovation.
All simulations were conducted on the \textit{Saguaro} cluster operated by the Fulton School of Engineering
at Arizona State University.
This research has made use of NASA's Astrophysics Data System.

\clearpage


\begin{thebibliography}{}

\bibitem[Ajiki et al.(2003)]{aji03} Ajiki, M., et al.\ 2003, 
\aj, 126, 2091 

\bibitem[Baldry et al.(2008)]{bal08} Baldry, I.~K., Glazebrook, K., \& 
Driver, S.~P.\ 2008, \mnras, 388, 945

\bibitem[Barton et al.(2004)]{ba04} Barton, E.~J., Dav{\'e}, R., Smith, J.-D.~T., Papovich, C., Hernquist, L., 
\& Springel, V.\ 2004, \apjl, 604, L1

\bibitem[Bianchi 
\& Schneider(2007)]{bia07} Bianchi, S., \& Schneider, R.\ 2007, \mnras, 378, 973 

\bibitem[Bond(2009)]{bon09} Bond, N.\ 2009, New Astronomy 
Review, 53, 42 

\bibitem[Cowie \& Hu(1998)]{cow98} Cowie, L.~L., \& Hu, E.~M.\ 1998, \aj, 115, 1319 


\bibitem[Crocce et al.(2006)]{cro06} Crocce, M., Pueblas, S., 
\& Scoccimarro, R.\ 2006, \mnras, 373, 369

\bibitem[Dav{\'e} et al.(2006)]{dav06} Dav{\'e}, R.,Finlator, K., \& Oppenheimer, B.~D.\ 2006, \mnras, 370, 273 

\bibitem[Davis et al.(1985)]{dav85} Davis, M., Efstathiou, 
G., Frenk, C.~S., \& White, S.~D.~M.\ 1985, \apj, 292, 371 

\bibitem[Dawson et al.(2004)]{da04} Dawson, S., et al.\  2004, \apj, 617, 707

\bibitem[Dawson et al.(2007)]{da07} Dawson, S., Rhoads, J.~E., Malhotra, S., Stern, D., Wang, J., Dey, A., Spinrad, H., \& Jannuzi, B.~T.\ 2007, \apj, 671, 1227


\bibitem[Dayal et al.(2008)]{day08} Dayal, P., Ferrara, A., 
\& Gallerani, S.\ 2008, \mnras, 389, 1683 


\bibitem[Dekel et al.(2009)]{dek09} Dekel, A., Sari, R., \& Ceverino, D.\ 2009, arXiv:0901.2458 

\bibitem[Dijkstra et al.(2006)]{dij06} Dijkstra, M., Haiman, 
Z., \& Spaans, M.\ 2006, \apj, 649, 14 

\bibitem[Dijkstra et al (2007)]{di07} Dijkstra, M., Wyithe, J. S. B., Haiman, Z.  2007, \mnras, 379, 253D


\bibitem[Fardal et al.(2001)]{far01} Fardal, M.~A., Katz, N., 
Gardner, J.~P., Hernquist, L., Weinberg, D.~H., 
\& Dav{\'e}, R.\ 2001, \apj, 562, 605 


\bibitem[Fernandez 
\& Komatsu(2008)]{fe08} Fernandez, E.~R., \& Komatsu, E.\ 2008, \mnras, 384, 1363 

\bibitem[Finkelstein et al.(2007)]{fi07} Finkelstein, S.~L., Rhoads, J.~E., Malhotra, S., Pirzkal, N., 
\& Wang, J.\ 2007, \apj, 660, 1023
 
 \bibitem[Finkelstein et al.(2008)]{fi08} Finkelstein, S.~L., 
Rhoads, J.~E., Malhotra, S., Grogin, N., \& Wang, J.\ 2008, \apj, 678, 655 

\bibitem[Fukugita et al.(1998)]{fuk98} Fukugita, M., Hogan, C.~J., \& Peebles, P.~J.~E.\ 1998, \apj, 503, 518 

\bibitem[Fynbo et al.(2001)]{fyn01} Fynbo, J.~U., M{\"o}ller, P., \& Thomsen, B.\ 2001, \aap, 374, 443

\bibitem[Gawiser et al.(2006)]{ga06} Gawiser, E., et al.\ 2006, \apjl, 642, L13

\bibitem[Gawiser et al.(2007)]{ga07} Gawiser, E., et al.\ 2007, \apj, 671, 278

\bibitem[Gawiser et al.(2007)]{ga07} Gawiser, E., et al.\ 2007, \apj, 671, 278

\bibitem[Genel et al.(2008)]{gen08} Genel, S., et al.\ 2008, 
\apj, 688, 789 

\bibitem[Gnedin et al.(2008)]{gne08} Gnedin, N.~Y., Kravtsov, 
A.~V., \& Chen, H.-W.\ 2008, \apj, 672, 765 

\bibitem[Gronwall et al.(2007)]{gro07} Gronwall, C., et al.\ 2007, \apj, 667, 79

\bibitem[Haiman 
\& Spaans(1999)]{ha99} Haiman, Z., \& Spaans, M.\ 1999, \apj, 518, 138 

\bibitem[Haiman et al.(2000)]{hai00} Haiman, Z., Spaans, M., 
\& Quataert, E.\ 2000, \apjl, 537, L5 

\bibitem[Haiman 
\& Cen(2005)]{hai05} Haiman, Z., \& Cen, R.\ 2005, \apj, 623, 627 

\bibitem[Hansen 
\& Oh(2006)]{ha06} Hansen, M., \& Oh, S.~P.\ 2006, \mnras, 367, 979 


\bibitem[Hinshaw et al.(2009)]{hin09} Hinshaw, G., et al. 2009, \apjs, 180, 225 

\bibitem[Hu et al.(1999)]{hu99} Hu, E.~M., McMahon, R.~G., 
\& Cowie, L.~L.\ 1999, \apjl, 522, L9 

\bibitem[Iliev et al.(2008)]{ili08} Iliev, I.~T., Shapiro, 
P.~R., McDonald, P., Mellema, G., \& Pen, U.-L.\ 2008, \mnras, 391, 63



\bibitem[Jimenez et al.(2005)]{jim05} Jimenez, R., Panter, 
B., Heavens, A.~F., \& Verde, L.\ 2005, \mnras, 356, 495 

\bibitem[Kashikawa et al.(2006)]{ka06} Kashikawa, N., et al.\ 2006, \apj, 648, 7 

 \bibitem[Kere{\v s} et al.(2009)]{ker09} Kere{\v s}, D., 
Katz, N., Fardal, M., Dav{\'e}, R., \& Weinberg, D.~H.\ 2009, \mnras, 395, 160

\bibitem[Kobayashi et al.(2007)]{kob07} Kobayashi, M.~A.~R., 
Totani, T., \& Nagashima, M.\ 2007, \apj, 670, 919

\bibitem[Kobayashi et al.(2009)]{kob09} Kobayashi, M.~A.~R., 
Totani, T., \& Nagashima, M.\ 2009, arXiv:0902.2882 

\bibitem[Kova{\v c} et al.(2007)]{ko07} Kova{\v c}, K., Somerville, R.~S., Rhoads, J.~E., Malhotra, S., 
\& Wang, J.\ 2007, \apj, 668, 15

\bibitem[Kudritzki et al.(2000)]{kud00} Kudritzki, R.-P., et 
al.\ 2000, \apj, 536, 19

 \bibitem[Landy 
\& Szalay(1993)]{la93} Landy, S.~D., \& Szalay, A.~S.\ 1993, \apj, 412, 64
 
\bibitem[Le Delliou et al.(2006)]{le06} Le Delliou, M., 
Lacey, C.~G., Baugh, C.~M., \& Morris, S.~L.\ 2006, \mnras, 365, 712

\bibitem[Leitherer et al.(1995)]{le95} Leitherer, C., 
Ferguson, H.~C., Heckman, T.~M., \& Lowenthal, J.~D.\ 1995, \apjl, 454, L19

\bibitem[Madau et al.(1998)]{mad98} Madau, P., Pozzetti, L., 
\& Dickinson, M.\ 1998, \apj, 498, 106 


\bibitem[Malhotra 
\& Rhoads(2002)]{mr02} Malhotra, S., \& Rhoads, J.~E.\ 2002, \apjl, 565, L71

\bibitem[Malhotra et al.(2003)]{ma03} Malhotra, S., Wang, J.~X., Rhoads, J.~E., Heckman, T.~M., 
\& Norman, C.~A.\ 2003, \apjl, 585, L25

\bibitem[Malhotra 
\& Rhoads(2004)]{mr04} Malhotra, S., \& Rhoads, J.~E.\ 2004, \apjl, 617, L5 


\bibitem[Mao et al.(2007)]{ma07} Mao, J., Lapi, A., Granato, 
G.~L., de Zotti, G., \& Danese, L.\ 2007, \apj, 667, 655 

\bibitem[Matsuda et al.(2005)]{mat05} Matsuda, Y., et al.\ 2005, \apjl, 634, L125


\bibitem[McQuinn et al.(2007)]{mcq07} McQuinn, M., Hernquist, 
L., Zaldarriaga, M., \& Dutta, S.\ 2007, \mnras, 381, 75

\bibitem[Mesinger \& Furlanetto(2008)]{mes08} Mesinger, A., \& Furlanetto, S.~R.\ 2008, \mnras, 386, 1990 

\bibitem[Murayama et al.(2007)]{mur07} Murayama, T., et al.\ 
2007, \apjs, 172, 523

\bibitem[Nagamine et al.(2008)]{na08} Nagamine, K., Ouchi, 
M., Springel, V., \& Hernquist, L.\ 2008, arXiv:0802.0228

\bibitem[Nagao et al.(2008)]{no08} Nagao, T., et al.\ 2008, \apj, 680, 100

\bibitem[Nilsson et 
al.(2007)]{nil07} Nilsson, K.~K., et al.\ 2007, \aap, 471, 71

\bibitem[Orsi et al.(2008)]{ors08} Orsi, A., Lacey, C.~G., 
Baugh, C.~M., \& Infante, L.\ 2008, \mnras, 391, 1589 

\bibitem[Ota et al.(2008)]{ota08} Ota, K., et al.\ 2008, 
\apj, 677, 12

\bibitem[Ouchi et al.(2008)]{ouc08} Ouchi, M., et al.\ 2008, 
\apjs, 176, 301

\bibitem[Ouchi et al.(2003)]{ouc03} Ouchi, M., et al.\ 2003, 
\apj, 582, 60

\bibitem[Partridge 
\& Peebles(1967)]{pa67} Partridge, R.~B., \& Peebles, P.~J.~E.\ 1967, \apj, 147, 868

\bibitem[Peebles(1980)]{peb80} Peebles, P.~J.~E.\ 1980, 
Research supported by the National Science Foundation.~Princeton, N.J., 
Princeton University Press, 1980.~435 p

\bibitem[Pentericci et al.(2009)]{pen09} Pentericci, L., Grazian, A., Fontana, A., Castellano, M., Giallongo, E., Salimbeni, S., \& Santini, P.\ 2009, \aap, 494, 553


\bibitem[Pirzkal et al.(2007)]{pi07} Pirzkal, N., Malhotra, 
S., Rhoads, J.~E., \& Xu, C.\ 2007, \apj, 667, 49

\bibitem[Razoumov 
\& Sommer-Larsen(2006)]{raz06} Razoumov, A.~O., \& Sommer-Larsen, J.\ 2006, \apjl, 651, L89

\bibitem[Rhoads et al.(2000)]{rh00} Rhoads, J.~E., Malhotra, 
S., Dey, A., Stern, D., Spinrad, H., 
\& Jannuzi, B.~T.\ 2000, \apjl, 545, L85

\bibitem[Rhoads 
\& Malhotra(2001)]{rho01} Rhoads, J.~E., \& Malhotra, S.\ 2001, \apjl, 563, L5 

\bibitem[Rhoads et al.(2004)]{rh04} Rhoads, J.~E., et al.\ 
2004, \apj, 611, 59 

\bibitem[Samui et al.(2009)]{sam09} Samui, S., Srianand, R., 
\& Subramanian, K.\ 2009, arXiv:0906.2312 

\bibitem[Scannapieco 
\& Thacker(2003)]{sca03} Scannapieco, E., \& Thacker, R.~J.\ 2003, \apjl, 590, L69 

\bibitem[Scannapieco 
\& Bildsten(2005)]{sca05} Scannapieco, E., \& Bildsten, L.\ 2005, \apjl, 629, L85 

\bibitem[Shapley et al.(2006)]{sha06} Shapley, A.~E., 
Steidel, C.~C., Pettini, M., Adelberger, K.~L., 
\& Erb, D.~K.\ 2006, \apj, 651, 688

\bibitem[Shimizu et al.(2007)]{shi07} Shimizu, I., Umemura, 
M., \& Yonehara, A.\ 2007, \mnras, 380, L49

\bibitem[Shioya et al.(2009)]{shio09} Shioya, Y., et al.\ 
2009, \apj, 696, 546

\bibitem[Shimasaku et al.(2006)]{shi06} Shimasaku, K., et 
al.\ 2006, \pasj, 58, 313 
	
\bibitem[Spergel et al.(2007)]{spe07} Spergel, D.~N., et al.\ 
2007, \apjs, 170, 377	 

\bibitem[Springel(2005)]{spr05} Springel, V.\ 2005, \mnras, 364, 1105 

\bibitem[Stark et al.(2007)]{st07} Stark, D.~P., Loeb, A., 
\& Ellis, R.~S.\ 2007, \apj, 668, 627


\bibitem[Stern et al.(2005)]{st05} Stern, D., Yost, S.~A., 
Eckart, M.~E., Harrison, F.~A., Helfand, D.~J., Djorgovski, S.~G., 
Malhotra, S., \& Rhoads, J.~E.\ 2005, \apj, 619, 12

\bibitem[Tapken et 
al.(2006)]{tap06} Tapken, C., et al.\ 2006, \aap, 455, 145 

\bibitem[Taniguchi et al.(2005)]{ta05} Taniguchi, Y., et 
al.\ 2005, \pasj, 57, 165 

\bibitem[Tasitsiomi(2006)]{tas06} Tasitsiomi, A.\ 2006, \apj, 
645, 792


\bibitem[Thacker \& Couchman (2006)]{tha06} Thacker, R. J. \& Couchman, H. M. P., 2006, Int. J. High Perf. Comp. \& Net., 4, 303

\bibitem[Venemans et 
al.(2005)]{ve05} Venemans, B.~P., et al.\ 2005, \aap, 431, 793

\bibitem[Wang et al.(2004)]{wa04} Wang, J.~X., et al.\ 2004, 
\apjl, 608, L21

\bibitem[Wang et al.(2005)]{wan05} Wang, J.~X., Malhotra, S., 
\& Rhoads, J.~E.\ 2005, \apjl, 622, L77 

\bibitem[Wang et al.(2009)]{wan09} Wang, J.-X., Malhotra, S., 
Rhoads, J.~E., Zhang, H.-T., \& Finkelstein, S.~L.\ 2009, arXiv:0907.0015 

\bibitem[Wood 
\& Loeb(2000)]{woo00} Wood, K., \& Loeb, A.\ 2000, \apj, 545, 86

\bibitem[Zehavi et al.(2005)]{ze05} Zehavi, I., et al.\ 
2005, \apj, 630, 1


\end{thebibliography}
\end{document}